\documentclass[twocolumn,prl,showpacs,superscriptaddress]{revtex4}

\usepackage{graphicx}
\usepackage{dcolumn}
\usepackage{bm}
\usepackage{braket}
\usepackage{leftindex}

\begin{document}
\title{Mapping the Second Landau Level PH-Pfaffian State to the Lowest Landau Level}
\author{Jian Yang}
\affiliation{Spinor Field LLC, Sugar Land, TX 77479, USA}

\begin{abstract}\
Recently we proposed a state described by the second Landau level (SLL) projection of the antiholomorphic Pfaffian wavefunction as a candidate for the ground state of the $\nu = 5/2$ fractional quantum Hall effect. In this paper we provide a rigorous mathematical proof that, when mapped to the lowest Landau level (LLL), the aforementioned  state, which we call the SLL PH-Pfaffian state, becomes exactly a LLL projected orbital angular momentum $l=-3$ pairing Pfaffian state, which we call the LLL ${\bf a}$ntiholomorphic ${\bf f}$-wave pairing Pfaffian state, or the LLL ${\bf af}$-Pfaffian state for short. We further prove that there is also an exact mapping between the upstream neutral Majorana fermion edge mode of the SLL PH-Pfaffian state and that of the LLL af-Pfaffian state. However, we find that while there exists an upstream neutral boson edge mode in the LLL af-Pfaffian state, it has no mapped counterpart in the SLL PH-Pfaffian state.

\end{abstract}
 \maketitle

\section {I. INTRODUCTION}
The leading candidates for the ground state of the $\nu=5/2$ fractional quantum Hall effect (FQHE) \cite{Willett} are the Pfaffian state\cite{MR}, the anti-Pfaffian state\cite{Levin}\cite{Lee} which is the particle-hole (PH) conjugate of the Pfaffian state, and the PH-Pfaffian state\cite{Son}\cite{Zucker}\cite{Yang}. There is a large number of numerical studies to support the Pfaffian state or the anti-Pfaffian state as a viable candidate for the ground state, but none for the PH-Pfaffian state.
The main reason that the PH-Pfaffian state has attracted considerable attention, despite of its scant numerical support, is because its edge structure supports a downstream charged boson edge mode and a counterpropagating upstream neutral Majorana fermion edge mode, making it the only state among the three leading candidates with the edge structure that is consistent with the experiments\cite{Banerjee}\cite{Dutta}. 

A possible resolution to the discrepancy between the experimental and the numerical results regarding the PH-Pfaffian state was offered recently\cite{Yang1} by a proposal of the following wave function 
\begin{equation}
\label{SPH-Pfaffian} {\Psi}_{SPH-Pf} = P_{SLL}{Pf} ( \frac{1}{z_i^*-z_j^* } ) \prod\limits_{i<j}^N (z_i-z_j)^2.
\end{equation}
where $z_j = x_j+iy_j$ is the complex coordinate of the $j_{th}$ electron, $N$ is the total number of electrons, and $Pf[A]$ is the Pfaffian of an antisymmetric matrix $A$, and $P_{SLL}$ is the second Landau level (SLL) projection operator. We call the state described by Eq.(\ref{SPH-Pfaffian}) the SLL PH-Pfaffian to distinguish it from the PH-Pfaffian state where the lowest Landau level (LLL) projection operator is used instead of $P_{SLL}$ in Eq.(\ref{SPH-Pfaffian})\cite{Zucker}\cite{Yang}. 
As shown in Ref.\cite{Yang1}, the SLL PH-Pfaffian represents a gapped, incompressible phase, and provides an excellent description for the exact ground state of finite-size SLL Coulomb-interacting electrons over a range of the short-distance interaction strength. Although the SLL PH-Pfaffian state is a SLL state and the anti-Pfaffian state is a LLL state, one can map the SPH-Pfaffian state to a LLL state, and the LLL mapped SPH-Pfaffian state 
is shown\cite{Yang1} to have a large overlap and similar low-energy orbital entanglement structures\cite{Li} with the anti-Pfaffian state.

In this paper, we will show that the SLL PH-Pfaffian state, when mapped to the LLL, will become exactly a LLL projected orbital angular momentum $l=-3$  pairing Pfaffian state. Because of their different analytic structures that dictate the corresponding low energy physics, the SLL PH-Pfaffian state and the LLL projected $l=-3$  pairing Pfaffian state can have different edge modes.

\section{II. EXACT MAPPING OF GROUND STATES}
\subsection{A. MAPPING}
We will begin with the LLL projected orbital angular momentum $l=-3$ pairing Pfaffian state described by the following wave function
\begin{equation}
\label{af-Pfaffian}{\Psi}_{af-Pf} =  P_{LLL}{Pf} ( \frac{1}{z_i^*-z_j^* } )^3 \prod\limits_{i<j}^N (z_i-z_j)^2
\end{equation}
where $P_{LLL}$ is the LLL projection operator. Because  it is a LLL projected ${\bf a}$ntiholomorphic ${\bf f}$-wave pairing Pfaffian state, we call it the LLL ${\bf af}$-Pfaffian state for short. It is interesting to note that in the framework of the Chern-Simons effective field theory\cite{Halperin1}\cite{Halperin2}, the anti-Pfaffian state would be described by the pairing in the $l=-3$ angular momentum channel, although there is no such a notion of the LLL projection in the effective field theory. 

Although ${\Psi}_{SPH-Pf}$ and ${\Psi}_{af-Pf}$ look differently, we have the following mathematical identity to map them exactly from each other:
\begin{equation}
\label{MainEq1} {\Psi}_{SPH-Pf} = (-2)^{N/2}(\prod\limits_{i = 1}^N a_i^\dagger) {\Psi}_{af-Pf}
\end{equation}
where  $a_i^\dagger$ is a Landau level raising operator, defined by $a_i^\dagger =  \sqrt{2}(-\partial_{z_i}+z_i^*/4)$
in the symmetric gauge ${\bf A} = \frac{B}{2}(y,-x)$. Together with the guiding center raising operator defined by $b_i^\dagger =  \sqrt{2}(\partial_{z_i^*}+z_i/4)$,
the single particle eigenstate can be generated as $\ket{nm}_i =  \frac{(a_i^\dagger)^n (b_i^\dagger)^m}{\sqrt{n!m!}}\ket{00}$
with energy $(n+\frac{1}{2})\hbar \omega_c$ and angular momentum $\hbar (m-n)$, where $\ket{00}$ is the vacuum state annihilated by $a_i$ and $b_i$, and $n$ and $m$ are non-negative integers. 

\subsection{B. PROOF}
In order to prove Eq.(\ref{MainEq1}), we first recognize that the  projection operator $P_{LLL}$ in Eq.(\ref{af-Pfaffian}) (or $P_{SLL}$ in Eq.(\ref{SPH-Pfaffian})) amounts to calculating the matrix elements of the Pfaffian factor ${Pf} ( \frac{1}{z_i^*-z_j^* } )^3$ (or ${Pf} \frac{1}{z_i^*-z_j^* }$) between the LLL (or SLL) $N$ particle basis and the LLL $N$ particle basis. Since the Pfaffian is a pair-wise function, we can project each pair separately, and the problem can be reduced to calculating the two-body matrix elements $\leftindex_j {\bra{1m_2'}}\leftindex_i{\bra{1m_1'}}\frac{1}{z_i^*-z_j^* } \ket{0m_1}_i\ket{0m_2}_j$
for $P_{SLL}$, and $\leftindex_j {\bra{0m_2'}}\leftindex_i{\bra{0m_1'}}\frac{1}{(z_i^*-z_j^*)^3 } \ket{0m_1}_i\ket{0m_2}_j$
for $P_{LLL}$, where $\ket{n_1m_1}_i\ket{n_2m_2}_j$ is a two-particle state for electron $i$ and electron $j$:
\begin{equation}
\label{TwoParticleState1} \ket{n_1m_1}_i \ket{n_2m_2}_j=  \frac{(a_i^\dagger)^{n_1}(b_i^\dagger)^{m_1} (a_j^\dagger)^{n_2} (b_j^\dagger)^{m_2}}{\sqrt{n_1!m_1!n_2!m_2!}}\ket{00}_i\ket{00}_j
\end{equation}
Because of the relative coordinates $z_i^*-z_j^*$ involved in the matrix element calculation, we transform the coordinates $z_i$ and $z_j$ to center of mass coordinate $z_c = (z_i+z_j)/2$ and relative coordinate $z_r = z_i-z_j$. Consequently, the raising operators ($a_i^\dagger$, $b_i^\dagger$) for particle $i$, and ($a_j^\dagger$, $b_j^\dagger$)  for particle $j$ are transformed to center of mass raising operators $a_c^\dagger = (a_i^\dagger+a_j^\dagger)/\sqrt{2}$, $b_c^\dagger = (b_i^\dagger+b_j^\dagger)/\sqrt{2}$, and relative raising operators $a_r^\dagger = (a_i^\dagger-a_j^\dagger)/\sqrt{2}$, $b_r^\dagger = (b_i^\dagger-b_j^\dagger)/\sqrt{2}$. As a result, an alternative representation for two-particle states can be constructed as follows
\begin{equation}
\label{TwoParticleState2} \ket{n_cm_c}_c\ket{n_rm_r}_r=  \frac{(a_c^\dagger)^{n_c}(b_c^\dagger)^{m_c} (a_r^\dagger)^{n_r} (b_r^\dagger)^{m_r}}{\sqrt{n_c!m_c!n_r!m_r!}}\ket{00}_c\ket{00}_r
\end{equation}
The states Eq.(\ref{TwoParticleState2}) are related to those in Eq.(\ref{TwoParticleState1})  by the following unitary transformation\cite{Sodemann}
\begin{equation}
\begin{aligned}
\label{UnitaryTransformationEq1}
 \ket{n_1m_1}_i \ket{n_2m_2}_j & =   \sum\limits_{\nu=0}^{n_1+n_2}\sum\limits_{\mu=0}^{m_1+m_2}R^{n_1+n_2}_{n_2,\nu}R^{m_1+m_2}_{m_2,\mu} \\
& \ket{n_1+n_2-\nu,m_1+m_2-\mu}_c\ket{\nu\mu}_r 
\end{aligned}
\end{equation}
where
\begin{equation}
\begin{aligned}
\label{UnitaryTransformationEq2}
R^{L}_{m,m'} = & \sqrt{\frac{\binom{L}{m}}{2^L \binom{L}{m'}}} \sum\limits_{\mu=max(0,m+m'-L)}^{min(m,m')} \\
& \binom{L-m}{m'-\mu} \binom{m}{\mu} (-1)^{\mu}
\end{aligned}
\end{equation}
The center of mass and relative basis of Eq.(\ref{TwoParticleState2}) is the most natural basis to calculate matrix elements $\leftindex_j {\bra{1m_2'}}\leftindex_i{\bra{1m_1'}}\frac{1}{z_i^*-z_j^* } \ket{0m_1}_i\ket{0m_2}_j$ and $\leftindex_j {\bra{0m_2'}}\leftindex_i{\bra{0m_1'}}\frac{1}{(z_i^*-z_j^*)^3 } \ket{0m_1}_i\ket{0m_2}_j$. The reason is that in the center of mass basis the matrix elements are diagonal, and while in the relative basis the matrix elements are given by
\begin{equation}
\begin{aligned}
\label{RelativeBasisMatrix}
& \leftindex_r {\bra{\nu'\mu'}}\frac{1}{(z_i^*-z_j^*)^l } \ket{0\mu}_r =  \leftindex_r {\bra{\nu'\mu'}}\frac{1}{{z_r^*}^l } \ket{0\mu}_r\\
& = (-1)^{\nu'}\sqrt{\frac{\mu!}{(\mu+\nu'+l)!\nu'!2^l}}\frac{\Gamma(\nu'+l)}{\Gamma(l)} \delta_{\mu',\mu+\nu'+l}
\end{aligned}
\end{equation}
In obtaining the above equation, we have used the following explicit form of the single particle eigenstate in the symmetric gauge
\begin{equation}
\label{SingleParticleES}
\braket{{\bf r}|\nu\mu}_r = (-1)^{\nu} \sqrt{\frac{{\nu}!}{2\pi 2^{\mu-\nu}{\mu}!}}z_r^{{\mu}-{\nu}}L^{{\mu}-{\nu}}_{{\nu}}(\frac{1}{2}|z_r|^2)e^{-\frac{1}{4}|z_r|^2}
\end{equation}
where $L^{{\mu}-{\nu}}_{{\nu}}$ is the associated Laguerre polynomial.
Using Eq.(\ref{UnitaryTransformationEq1}),  Eq.(\ref{UnitaryTransformationEq2}), Eq.(\ref{RelativeBasisMatrix}), noticing $R^{0}_{0,0} = 1$ and $R^{2}_{1,2} = -1/\sqrt{2}$, it is straightforward to establish the following relationship between the two-body matrix elements $\leftindex_j {\bra{1m_2'}}\leftindex_i{\bra{1m_1'}}\frac{1}{z_i^*-z_j^* } \ket{0m_1}_i\ket{0m_2}_j$ and $\leftindex_j {\bra{0m_2'}}\leftindex_i{\bra{0m_1'}}\frac{1}{(z_i^*-z_j^*)^3 } \ket{0m_1}_i\ket{0m_2}_j$
\begin{widetext}
\begin{equation}
\begin{aligned}
\label{MappingEq1}
& \leftindex_j {\bra{1m_2'}}\leftindex_i{\bra{1m_1'}}\frac{1}{z_i^*-z_j^* } \ket{0m_1}_i\ket{0m_2}_j  = -2\leftindex_j {\bra{0m_2'}}\leftindex_i{\bra{0m_1'}}\frac{1}{(z_i^*-z_j^*)^3 } \ket{0m_1}_i\ket{0m_2}_j \\
& = -\sum\limits_{\mu=0}^{m_1+m_2}R^{m_1+m_2}_{m_2,\mu}R^{m_1'+m_2'}_{m_2',\mu+3} 
\sqrt{\frac{\mu!}{2(\mu+3)!}}\delta_{m_1'+m_2',m_1+m_2+3}
\end{aligned}
\end{equation}
\end{widetext}
Using Eq.(\ref{MappingEq1}) and the fact $\ket{1m}_i = a_i^\dagger \ket{0m}_i$, the mapping equation Eq.(\ref{MainEq1}) follows readily.

\section{III. EXACT MAPPING OF EDGE MODES}
In this section, we will focus only on the upstream neutral edge modes and will make a brief  remark on the downstream charged boson charge mode in the next section as it is fairly trivial.

\subsection{A. EDGE MODES OF SLL PH-PFAFFIAN}
Following similar arguments regarding the edge mode constructions of the Pfaffian state\cite{Milovanovic}, the SLL PH-Pfaffian state has an upstream neutral Majorana edge mode described by:
\begin{widetext}
\begin{eqnarray}
\label{SPH-PfaffianMajorana}{\Psi}_{SPH-Pf}^{m_1, m_2, \cdot\cdot\cdot, m_{F}}(z_1, z_2, \cdot\cdot\cdot, z_N) =  P_{SLL}
\hat{A}  ( \prod\limits_{k=1}^{F} {z_k^*}^{m_k}
\prod\limits_{l=1}^{(N-F)/2} \frac{1}{z_{F+2l-1}^*-z_{F+2l}^*} )\prod\limits_{i<j}^N (z_i-z_j)^2
\end{eqnarray}
\end{widetext}
where $\hat{A}$ is an anti-symmetrization operator with respect to the electron coordinates $z_1, z_2, \cdot\cdot\cdot, z_N$, $F$ is an integer $F \leq N$, and $m_k$ can be taken ordered and distinct non-negative integers due to the antisymmetrization: $0 \leq m_1 < m_2 < \cdot\cdot\cdot < m_{F-1} < m_F$. The wavefunction Eq.(\ref{SPH-PfaffianMajorana}) describes $(N-F)/2$ paired electrons, with $F$ fermions left unpaired resulting from $F/2 $ broken pairs (for the sake of simplicity we assume $F$ is even). The $F$ unpaird fermions represent upstream neutral Majorana fermion edge mode occupying single particle wave functions ${z^*}^{m_k}$. The angular momentum relative to the ground state is $\Delta{M} = M-M_0 = -\prod\limits_{k=1}^{F}(m_k+\frac{1}{2})$, where $M = -\prod\limits_{k=1}^{F}m_k + N(N-1)+(N-F)/2 $ for ${\Psi}_{SPH-Pf}^{m_1, m_2, \cdot\cdot\cdot, m_{F}}(z_1, z_2, \cdot\cdot\cdot, z_N)$ in Eq.(\ref{SPH-PfaffianMajorana}) and $M_0 = N(N-1)+N/2 $ for ${\Psi}_{SPH-Pf}$ in  Eq.(\ref{SPH-Pfaffian}).

\subsection{B. EDGE MODES OF LLL af-PFAFFIAN}
Following the similar argument, the LLL af-Pfaffian state has an upstream neutral Majorana fermion edge mode described by:
\begin{widetext}
\begin{eqnarray}
\label{af-PfaffianMajorana}
{\Psi}_{af-Pf}^{m_1, m_2, \cdot\cdot\cdot, m_{F}}(z_1, z_2, \cdot\cdot\cdot, z_N) =  P_{LLL}
\hat{A}  ( \prod\limits_{k=1}^{F} {z_k^*}^{m_k}
\prod\limits_{l=1}^{(N-F)/2} \frac{1}{(z_{F+2l-1}^*-z_{F+2l}^*)^3} )\prod\limits_{i<j}^N (z_i-z_j)^2
\end{eqnarray}
\end{widetext}
The angular momentum  of state Eq.(\ref{af-PfaffianMajorana}) is $M = -\prod\limits_{k=1}^{F}m_k + N(N-1)+3(N-F)/2 $. Since the angular momentum  of state Eq.(\ref{af-Pfaffian}) is $M_0 = N(N-1)+3N/2 $, the angular momentum relative to the ground state is $\Delta{M} = M-M_0 = -\prod\limits_{k=1}^{F}(m_k+\frac{3}{2})$.

The LLL af-Pfaffian state has also an upstream neutral boson edge mode. Unlike the upstream Majorana edge mode which is formed by breaking up electron pairs, the boson mode is formed by changing the angular momentum of pairs from $l = -3$ to $l = -1$:
\begin{widetext}
\begin{eqnarray}
\label{af-PfaffianBoson}
{\Psi}_{af-Pf}^{m_1, m_2, \cdot\cdot\cdot, m_{B}}(z_1, z_2, \cdot\cdot\cdot, z_N) =  P_{LLL}
\hat{A}  ( \prod\limits_{k=1}^{B} {z_k^*}^{m_k} \prod\limits_{k=1}^{B/2}  \frac{1}{z_{2k-1}^*-z_{2k}^*}
\prod\limits_{l=1}^{(N-B)/2} \frac{1}{(z_{F+2l-1}^*-z_{F+2l}^*)^3} )\prod\limits_{i<j}^N (z_i-z_j)^2
\end{eqnarray}
\end{widetext}
where $B$ is an even integer $B \leq N$. The wavefunction Eq.(\ref{af-PfaffianBoson}) represent $(N-B)/2$ $l = -3$ paired electrons, with $B/2$ paired electrons changing the angular momentum of pairs from $l = -3$ to $l = -1$. Because $l = -1$ pairing is still in place for the $B/2$ paired electrons, the integers $m_k$ can be taken ordered but not necessarily  distinct, $0 \leq m_1 \leq m_2 \leq \cdot\cdot\cdot \leq m_{B-1} \leq m_B$, which therefore describe $B$ bosons representing upstream neutral boson edge mode occupying single particle wavefunctions ${z^*}^{m_k}$. 
The angular momentum  of state Eq.(\ref{af-PfaffianBoson}) is $M = -\prod\limits_{k=1}^{B}m_k + N(N-1)+B/2+3(N-B)/2 $. Since the angular momentum  of state Eq.(\ref{af-Pfaffian}) is $M_0 = N(N-1)+3N/2 $, the angular momentum of ${\Psi}_{af-Pf}^{m_1, m_2, \cdot\cdot\cdot, m_{B}}(z_1, z_2, \cdot\cdot\cdot, z_N)$ relative to the ground state is $\Delta{M} = M-M_0 = -\prod\limits_{k=1}^{B}(m_k+1)$.

One can also combine Eq.(\ref{af-PfaffianMajorana}) and Eq.(\ref{af-PfaffianBoson}) to generate a state with both upstream neutral Majorana edge mode and boson edge mode present simultaneously.

\subsection{C. MAPPING OF MAJORANA EDGE MODES}

Using Eq.(\ref{MappingEq1}), together with the following equation:
\begin{equation}
\label{MappingEq2}
{\bra{1m'}}{z^*}^{l+1}\ket{0m} = \sqrt{2} (l+1) {\bra{0m'}}{z^*}^l\ket{0m}
\end{equation}
one can map the upstream neutral Majorana mode of the SLL PH-Pfaffian state Eq.(\ref{SPH-PfaffianMajorana}) and that of the LLL af-Pfaffian state Eq.(\ref{af-PfaffianMajorana}) as follows:
\begin{widetext}
\begin{equation}
\label{MainEq2}
{\Psi}_{SPH-Pf}^{m_1+1, m_2+1, \cdot\cdot\cdot, m_{F}+1}(z_1, z_2, \cdot\cdot\cdot, z_N) = (-2)^{N/2}\prod\limits_{i = 1}^F(m_i+1)(\prod\limits_{i = 1}^N a_i^\dagger){\Psi}_{af-Pf}^{m_1, m_2, \cdot\cdot\cdot, m_{F}}(z_1, z_2, \cdot\cdot\cdot, z_N)
\end{equation}
\end{widetext}
Although, as shown in Eq.(\ref{MainEq2}), the edge mode quantum number $m_i$ is increased by $1$ when mapping from LLL af-Pfaffian state to the SLL PH-Pfaffian state, the physical angular momentum, which is $m_i-n_i$, actually remains the same because LLL af-Pfaffian state is a LLL state ($n_i = 0$) while SLL PH-Pfaffian state is a SLL state ($n_i = 1$). This increase of $m_i$ by $1$ also makes the angular momentum relative to the ground state on both sides of Eq.(\ref{MainEq2}) the same, $\Delta{M} = M-M_0 = -\prod\limits_{k=1}^{F}(m_k+\frac{3}{2})$. 
Furthermore, Eq.(\ref{MainEq2}) shows that ${\Psi}_{SPH-Pf}^{m_1, m_2, \cdot\cdot\cdot, m_{F}}(z_1, z_2, \cdot\cdot\cdot, z_N)$ is non-vanishing only when 
$1 \leq m_1 < m_2 < \cdot\cdot\cdot m_{F-1} < m_F$ instead of $0 \leq m_1 < m_2 < \cdot\cdot\cdot m_{F-1} < m_F$.

\subsection{D. NO MAPPING OF BOSON EDGE MODE}
To see if one can map the upstream neutral boson edge mode of the LLL af-Pfaffian state Eq.(\ref{af-PfaffianBoson}) to any of the SLL PH-Pfaffian state, we need to calculate the following matrix $\leftindex_j {\bra{0m_2'}}\leftindex_i{\bra{0m_1'}}\frac{{z_i^*}^{l_i}{z_j^*}^{l_j}}{z_i^*-z_j^* } \ket{0m_1}_i\ket{0m_2}_j$
with $l_i$ and $l_j$ being non-negative integers. Since ${z_i^*}^{l_i}{z_j^*}^{l_j}$ can always be written in terms of the center of mass coordinates $z_c = (z_i+z_j)/2$ and relative coordinate $z_r = z_i-z_j$,
${z_{i}^*}^{l_{i}}{z_{j}^*}^{l_{j}} =  \sum\limits_{k_i=0}^{l_i}\sum\limits_{k_j=0}^{l_j}(-1)^{k_j}(1/2)^{k_i+k_j}\binom{l_i}{k_i}\binom{l_j}{k_j} (z_r^*)^{k_i+k_j} (z_c^*)^{l_i+l_j-k_i-k_j}$,
it is suffice to calculate the following matrix $\leftindex_j {\bra{0m_2'}}\leftindex_i{\bra{0m_1'}}\frac{(z_c^*)^{l'}}{(z_r^*)^l} \ket{0m_1}_i\ket{0m_2}_j$
with $l$ and $l'$ being any integers. Again we will use the center of mass and relative basis of Eq.(\ref{TwoParticleState2}) to calculate matrix elements. The difference now is the matrix elements are no longer diagonal in the center of mass basis and needs to be calculated  in the same way as for the relative basis.

Using Eq.(\ref{UnitaryTransformationEq1}), Eq.(\ref{UnitaryTransformationEq2}) (along with some useful results $R^{0}_{0,0} = 1$,  $R^{2}_{1,1} = 0$, and $R^{2}_{1,2} = -1/\sqrt{2}$), Eq.(\ref{SingleParticleES}), and the same form of Eq.(\ref{SingleParticleES}) for the single particle eigenstate in the center of mass basis by substituting $z_c$ for $z_r$, we obtain the following equation
\begin{widetext}
\begin{equation}
\begin{aligned}
\label{MappingEq3}
& \leftindex_j {\bra{1m_2'}}\leftindex_i{\bra{1m_1'}}\frac{(z_i^*+z_j^*)^{L+2}}{z_i^*-z_j^*}  \ket{0m_1}_i\ket{0m_2}_j = 
-2 \leftindex_j {\bra{0m_2'}}\leftindex_i{\bra{0m_1'}}\frac{(z_i^*+z_j^*)^{L+2}}{(z_i^*-z_j^*)^3}  \ket{0m_1}_i\ket{0m_2}_j \\
&+ 2(L+1)(L+2) \leftindex_j {\bra{0m_2'}}\leftindex_i{\bra{0m_1'}}\frac{(z_i^*+z_j^*)^{L}}{z_i^*-z_j^*}  \ket{0m_1}_i\ket{0m_2}_j 
\end{aligned}
\end{equation}
\end{widetext}
where  the matrix elements $\leftindex_j {\bra{0m_2'}}\leftindex_i{\bra{0m_1'}}\frac{{(z_i^*+z_j^*)}^{l'}}{{(z_i^*-z_j^*)}^l}  \ket{0m_1}_i\ket{0m_2}_j$ can be calculated explcitly
\begin{widetext}
\begin{equation}
\begin{aligned}
& \leftindex_j {\bra{0m_2'}}\leftindex_i{\bra{0m_1'}}\frac{{(z_i^*+z_j^*)}^{l'}}{{(z_i^*-z_j^*)}^l}  \ket{0m_1}_i\ket{0m_2}_j = \\
& \sum\limits_{\mu=0}^{m_1+m_2}R^{m_1+m_2}_{m_2,\mu}R^{m_1'+m_2'}_{m_2',\mu+l} 
\sqrt{\frac{2^{3l'-l}\mu!(m_1+m_2-\mu)!}{(\mu+l)!(m_1+m_2-\mu-l')!}}\delta_{m_1'+m_2',m_1+m_2+l-l'}
\end{aligned}
\end{equation}
\end{widetext}
It is noted that Eq.(\ref{MappingEq1}) is a special case of Eq.(\ref{MappingEq3}) with $L = -2$. 

An immediate conclusion can be drawn from Eq.(\ref{MappingEq3}) is: The boson edge mode of the LLL af-Pfaffian state, as represented by the second term with non-negative $L$ on the right side of Eq.(\ref{MappingEq3}), by itself, cannot be mapped to any low energy excitations of the SLL PH-Pfaffian state. In other words, the upstream neutral boson edge mode of the LLL af-Pfaffian state has no mapped counterpart in the SLL PH-Pfaffian state, it is simply mapped out of the low energy Hilbert space of the SLL PH-Pfaffian state. Another observation, except for the second term on the right side of Eq.(\ref{MappingEq3}), the other two terms with non-negative $L$ in Eq.(\ref{MappingEq3}) are necessarily representing bulk excitations (quasielectrons).

\section{IV. FINAL REMARKS}
Before closing, we would like to make two final remarks. The first one is on a rather trivial matter which we have ignored so far - the downstream charged boson edge mode. Since it is generated by multiplying symmetric holomorphic polynomials $\sum\limits_{i}^N z_i^n$ with $n$ being a non-negative integer to the Jastrow factor $\prod\limits_{i<j}^N (z_i-z_j)^2$ in Eq.(\ref{SPH-Pfaffian}) and in  Eq.(\ref{af-Pfaffian}),  the mapping equation between the downstream charged boson edge mode of SLL PH-Pfaffian and that of the LLL af-Pfaffian will be exactly the same as Eq.(\ref{MainEq1}), only recognizing that the two-particle state $\ket{0m_1}_i\ket{0m_2}_j$ in calculating matrix elements $\leftindex_j {\bra{1m_2'}}\leftindex_i{\bra{1m_1'}}\frac{1}{z_i^*-z_j^* } \ket{0m_1}_i\ket{0m_2}_j$, or in $\leftindex_j {\bra{0m_2'}}\leftindex_i{\bra{0m_1'}}\frac{1}{(z_i^*-z_j^*)^3 } \ket{0m_1}_i\ket{0m_2}_j$, would represent $\sum\limits_{i}^N z_i^n \prod\limits_{i<j}^N (z_i-z_j)^2$ instead of $\prod\limits_{i<j}^N (z_i-z_j)^2$. Secondly, the mapping between the SLL PH-Pfaffian state and LLL af-Pfaffian state is proven mathematically exact in the disk geometry in this paper. However we find it is not exact in the spherical geometry\cite{Haldane}, although the overlap between the two is found to be remarkably large for finite systems. For example, the overlap (the square root of the inner product of the wave functions) between the LLL mapped SLL PH-Pfaffian state and LLL af-Pfaffian state on a sphere is $0.9980$ for $6$ electrons, and $0.9977$ for $8$ electrons. A detailed study on the subject in the spherical geometry will be reported elsewhere.

\end{document}